\documentclass[conference]{IEEEtran}
\IEEEoverridecommandlockouts
\usepackage{cite}
\usepackage{amsmath,amssymb,amsfonts}
\usepackage{algorithmic}
\usepackage{graphicx}
\usepackage{textcomp}
\usepackage{xcolor}
\usepackage{array}
\usepackage{float}
\def\BibTeX{{\rm B\kern-.05em{\sc i\kern-.025em b}\kern-.08em
    T\kern-.1667em\lower.7ex\hbox{E}\kern-.125emX}}
\begin{document}

\title{r/ecommender - A Personalized Subreddit Recommendation Engine\\}

\author{
\IEEEauthorblockN{Abhishek K Das}
\IEEEauthorblockA{Department of Computer Science\\
Texas A\&M University\\
College Station, Texas\\
Email: abkds@tamu.edu}\\   
\IEEEauthorblockN{Janvi Palan}
\IEEEauthorblockA{Department of Computer Science\\
Texas A\&M University\\
College Station, Texas\\
Email: janvipalan@tamu.edu}
\and
\IEEEauthorblockN{Nikhil Bhat}
\IEEEauthorblockA{Department of Computer Science\\
Texas A\&M University\\
College Station, Texas\\
Email: nbhat510@tamu.edu}\\              
\IEEEauthorblockN{Sukanto Guha}
\IEEEauthorblockA{Department of Computer Science\\
Texas A\&M University\\
College Station, Texas\\
Email: sukantoguha@tamu.edu}
}
\maketitle

\begin{abstract}
This project aims to improve upon the generic recommendations that Reddit provides for its users. We propose a novel personalized recommender system that learns from both, the presence and the content of user-subreddit interaction, using implicit and explicit signals to provide robust recommendations.
\end{abstract}


\section{Introduction}
Reddit, with more than 540 Million Active users and 138K+ subreddits, is the third most frequently visited website in the US. The website comprises of more than a million 'subreddits', wherein each subreddit caters to a specific topic. For example, the subreddit r/soccer is the subreddit for the game of soccer. Each subreddit has between a handful to millions of users who post content and comment on the posted content, and these comments are what form the aggregated dataset for our work.

While it's tagline is 'The front page of the internet', it still lacks in its subreddit recommendation system, which usually provides generic recommendations for users. While websites such as Amazon and LinkedIn have mastered the art of making personalized recommendations based on user profiles, browsing and interaction history, Reddit still relies on trending topics and moderators to generate a list of recommendations. Moreover, these are not unique or personalized to the user's taste. 

Our project proposes a novel model that takes in implicit and explicit factors from the interactions that a user has had with a subreddit, and tackles the problem of generating personalized recommendations in various ways. We use methods from traditional Information Retrieval and Natural Language Processing, to test various models and evaluate them on the basis of their performance. We propose a hybrid recommender system, which relies on both, collaborative filtering and content-based features, to build a system that can tackle the problem of cold start as well as low computation processing capabilities.

\section{Related Work} 
There have been previous models which have worked in the domain to provide and improve subreddit recommendations. Akash Japi, in his \cite{b6} blog article wrote a web crawler to scrape user comments belonging to users found on top comments from the subreddit r/all which itself aggregates the top posts of reddit. He then created vectors for each users and then used k-NN measures to find similar users.

Furthermore, although in a different domain, we found the work presented by He and McAuley in VBPR: Visual Bayesian Personalized Ranking from Implicit Feedback\cite{b2} quite relevant. The authors use a visual embedding of items to suggest better recommendations by grouping visually similar items together. Based on this system, we develop a \textit{Textual Bayesian Personalized Ranking} recommendation system, which creates lower space embeddings for both users and subreddits to suggest similar subreddits.

\section{Methodology}

\subsection{Dataset}\label{AA}
 
 We obtained a dataset which comprises of user comments on Reddit from the month of January 2015. It contains 57 million comments from Reddit users. Since Reddit does not release the user-subreddit subscription data, implicit and explicit indicators of user interest have to be derived from this dataset. We further processed this dataset to improve the quality by performing the following steps.
 \begin{itemize}
 \item Removing user - subreddit interactions which were lesser than 30 characters and users with fewer than 5 comments
 \item Removing bots and [deleted] comments
 \item Dividing the dataset into two parts - dataset A with interactions, and dataset B with comments as well as interactions
 
 \end{itemize}
Our final datasets consisted of \textbf{28 million comments, 735834 users and 14842 subreddits}. All the comments were filtered to lowercase, and stopwords and punctuation were removed. 

\begin{figure}[htbp]
\centerline{\includegraphics[scale=0.60]{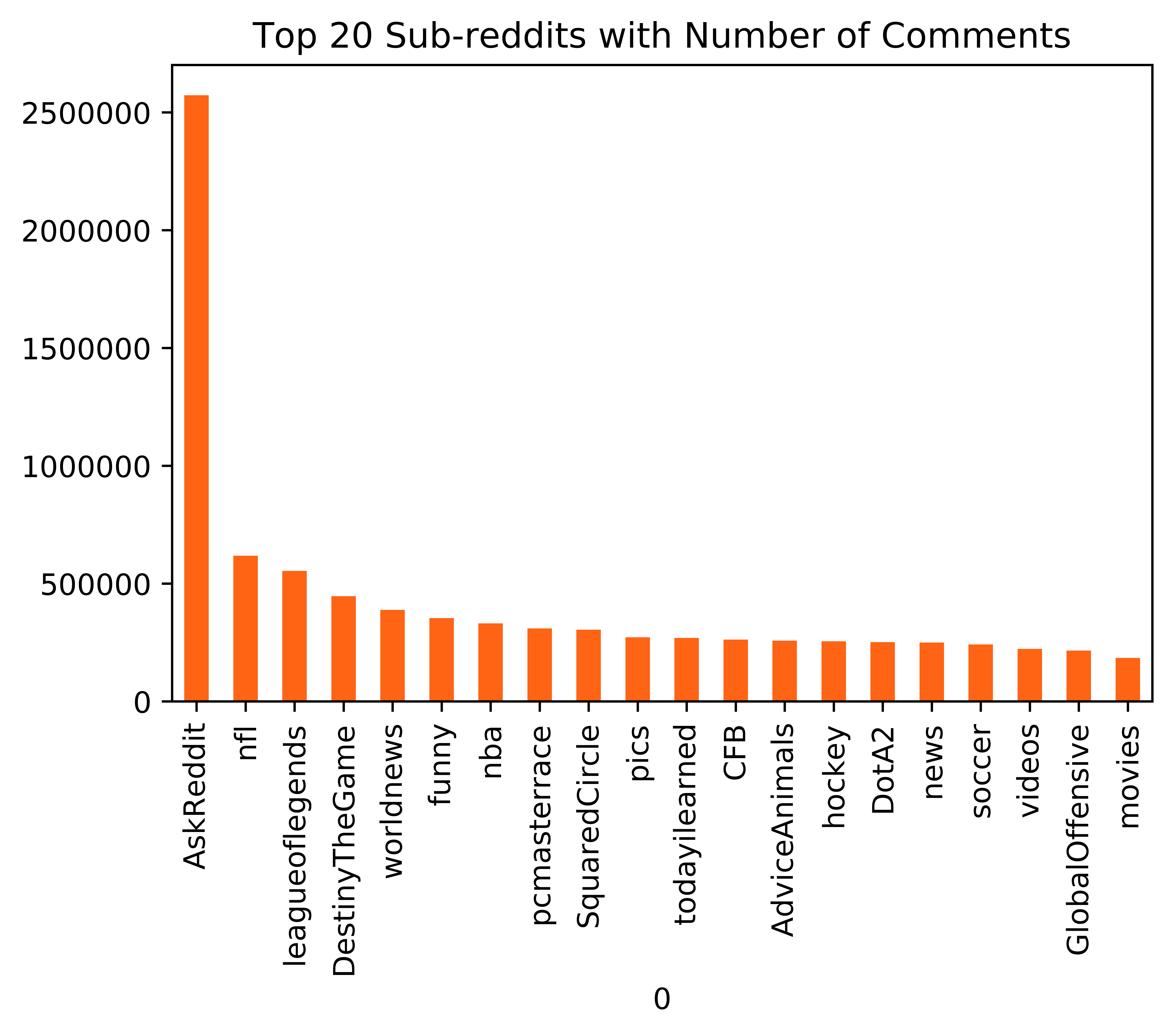}}
\caption{Most popular subreddits by comments in January 2015}
\label{subredditspop}
\end{figure}

The final dataset contained subreddits of varying popularity. One aim of our project is to try to not simply recommend the most popular, or most interacted with subreddits. We explore the dataset to understand what fraction of the data was dominated by subreddits. This is shown in Figure \ref{subredditspop}, which indicates that nearly 20\% of the comments in our dataset were made on the top 30 subreddits.  

\subsection{Matrix factorization - ALS }

To create a baseline for our recommender system, and to leverage the large amount of informative interactions we have between users and subreddits, we use the dataset A and implement a user-subreddit collaborative filtering based model, that uses Matrix Factorization using Alternating Least Squares (ALS) as the optimization function[cite]. For this model, we assume the presence of an interaction between a user and a subreddit as a positive rating, assigning it a rating of 1, and the absence is assigned a rating of 0.

This information is then used to minimize the objective function as described in [Implicit cf paper cite], which calculates the user factors and the subreddit factors by projecting them into a common latent factor space where they can be directly compared. 

\subsection{Bayesian Personalized Ranking - BPR }
\begin{figure}[htbp]
\centerline{\includegraphics[scale=0.60]{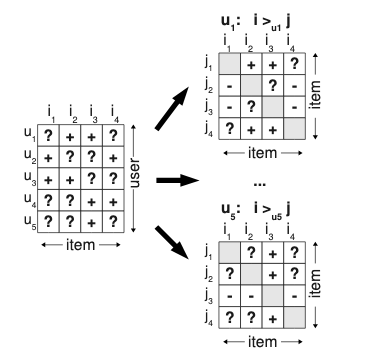}}
\caption{Bayesian Personalized Rankings}
\label{BPR}
\end{figure}
The Bayesian Personalized Ranking works with the goal to estimate a personalized ranking function for each user with respect to all pairs of items, $i$ and $j$. In this paper\cite{b1}, the authors use implicit signals to compare pairs of items for users to create rankings as shown below in Figure \ref{BPR}.  For each user $u$, the value $x_{u,i,j}$ is calculated. This value is positive is user $u$ prefers item $i$ over item $j$, and negative if vice-versa. This value is calculated by any method that can approximate the relationship of a user with any two items. In our experiments, we estimate this value as follows:
$$\hat{x}_{u,i,j} = \gamma_{u}*\gamma_{i} - \gamma_{u}*\gamma_{j}$$

where the $\gamma$ terms are Matrix Factorization based latent factors.

\subsection{Textual Bayesian Personalized Ranking (t-BPR)}

The concept of t-BPR takes inspiration from v-BPR [cite paper] where we combine the knowledge of implicit preferences for a certain subreddit for a user with the actual explicit features, that is, the content of the comment. Similar to visual factors in vBPR, we used textual factors for both, the items (in this case, subreddits), and the users, and incorporated this in our algorithm. We trained the textual factors for both - \textbf{Subreddit2Vec} and \textbf{User2Vec}, using the Doc2Vec algorithm [cite paper]. We then use these factors in our textual BPR algorithm using two methods: \textbf{vanilla t-BPR} and \textbf{Learnt t-BPR}. 
\begin{figure}[htbp]
\centerline{\includegraphics[scale=0.5]{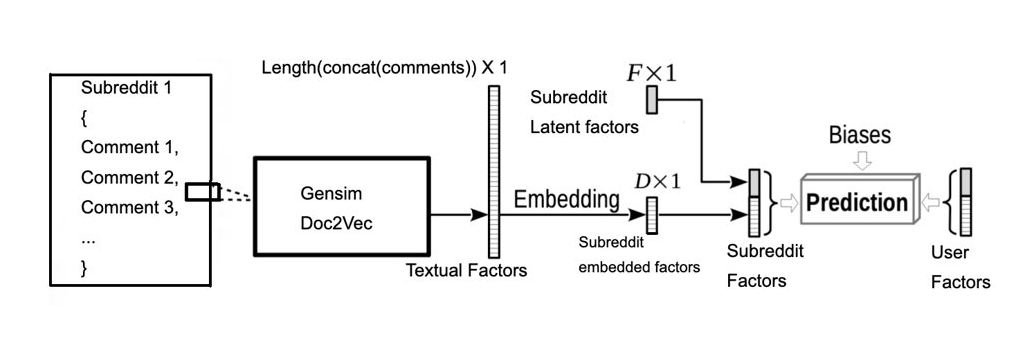}}
\caption{Textual BPR architecture}
\label{tbpr}
\end{figure}
\subsubsection{Vanilla t-BPR}
 
Using the independently learnt textual factors for each subreddit and each user, we calculated the values for $x_{u,i,j}$ for all pairs of subreddits $i$ and $j$, for each user. This is done by calculating the preference of any user $u$ for any subreddit $i$ using the following equation:
$$\hat{x}_{u,i} = \alpha + \beta_{u} + \beta_{i} + \gamma_{u}^{T}\gamma_{i} + \theta_{u}^{T}\theta_{i}$$

where $\alpha$ is the global bias, $\beta_{u}$ is the bias for user u, $\beta_{i}$ is the bias for subreddit i, and the $\gamma$ values are the subreddit2vec and user2vec values for that particular subreddit and user.  $\theta_{u}$ and $\theta_{i}$ are the newly introduced textual factors whose inner product models the interaction between the users and the subreddits in the form of comments' representations in $D$ dimensions. 

The calculated estimated values for $x$ are then evaluated using the AUC methodology described in the Evaluation section. 


\subsubsection{Learnt t-BPR}

We further work on our t-BPR to learn the user embedding kernel which linearly transforms the high-dimensional features of the users into a much lower-dimensional ‘textual rating’ space. This is represented by the following:
$$ \theta_{i} = Ef_{i}$$

The final prediction model is:
$$\hat{x}_{u,i} = \alpha + \beta_{u} + \beta_{i} + \gamma_{u}^{T}\gamma_{i} + \theta_{u}^{T}Ef_{i} + \beta^{'T}f_{i} $$

The following optimization criterion is used for personalized ranking:
$$ \sum_{(u,i,j) \in D_{S}} ln\sigma(x_{uij}) - \lambda_{\Theta}||\Theta||^{2}$$ 
where $\sigma$ is the logistic (sigmoid) function and $\lambda_{\Theta}$ is a model-specific regularization hyperparameter.

\section*{Evaluation}

For evaluation, we split our data into two sets- training and test data. We initially have a list of subreddits a user subscribes to. We take out 10\% of subreddits associated with a user and add them to our test set. We used Stratified Sampling to split the dataset, in order to ensure that there is enough training data for each user. Once training is complete, we test how many of the subreddits we removed in the initial set are present in our recommendations. 

We use the Area-Under-the-Curve (AUC) as a metric to quantify our evaluation. We considered various evaluation criteria, and decided that Area Under ROC Curve(AUC) as the best for our recommender system\cite{b3}.  This evaluation criteria gives the probability that a classifier will rank a randomly chosen positive instance higher than a randomly chosen negative one. Comparing pairs of subreddits to see their relative ranking is the most important feature in a subreddit recommender system, so measuring AUC was the best fit for evaluation. AUC is a well-defined criterion because our model is comparison-based, without any ground-truth for our recommender system. 

We take the definition as given in the [cite BPR paper] BPR paper. AUC is defined as:
$$AUC = \frac{1}{| U |} \sum_u \frac{1}{|E(u)|} \sum_{(i, j) \in E(u)} \delta(\hat{x}_{ui} - \hat{x}_{uj}) $$

where $$E(u) := \{(i, j) | (u, i) \in S_{test} ∧ (u, j) \notin (S_{test} ∪ S_{train})\}$$

\section*{Results}

\begin{figure}[htbp]
\centerline{\includegraphics[scale=0.2]{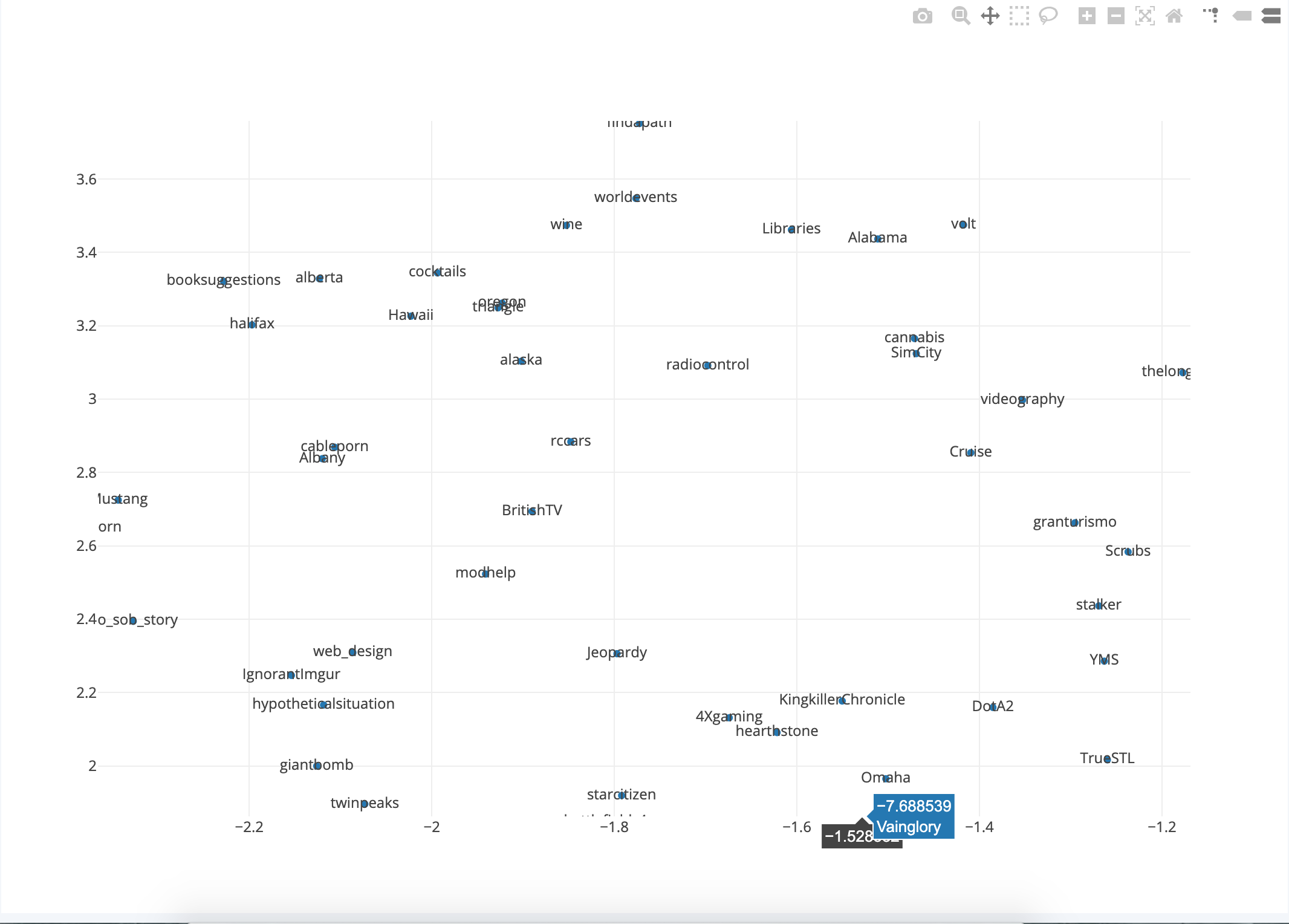}}
\caption{t-SNE plot for subreddits}
\label{tsne}
\end{figure}

We plotted t-SNE plots\cite{b4} for our subreddits based on the Subreddit2Vec vectors that we generated to observe the similarities between subreddits and find interesting combinations between the as shown in Figure \ref{tsne}.

\begin{figure}[htbp]
\centerline{\includegraphics[scale=0.6]{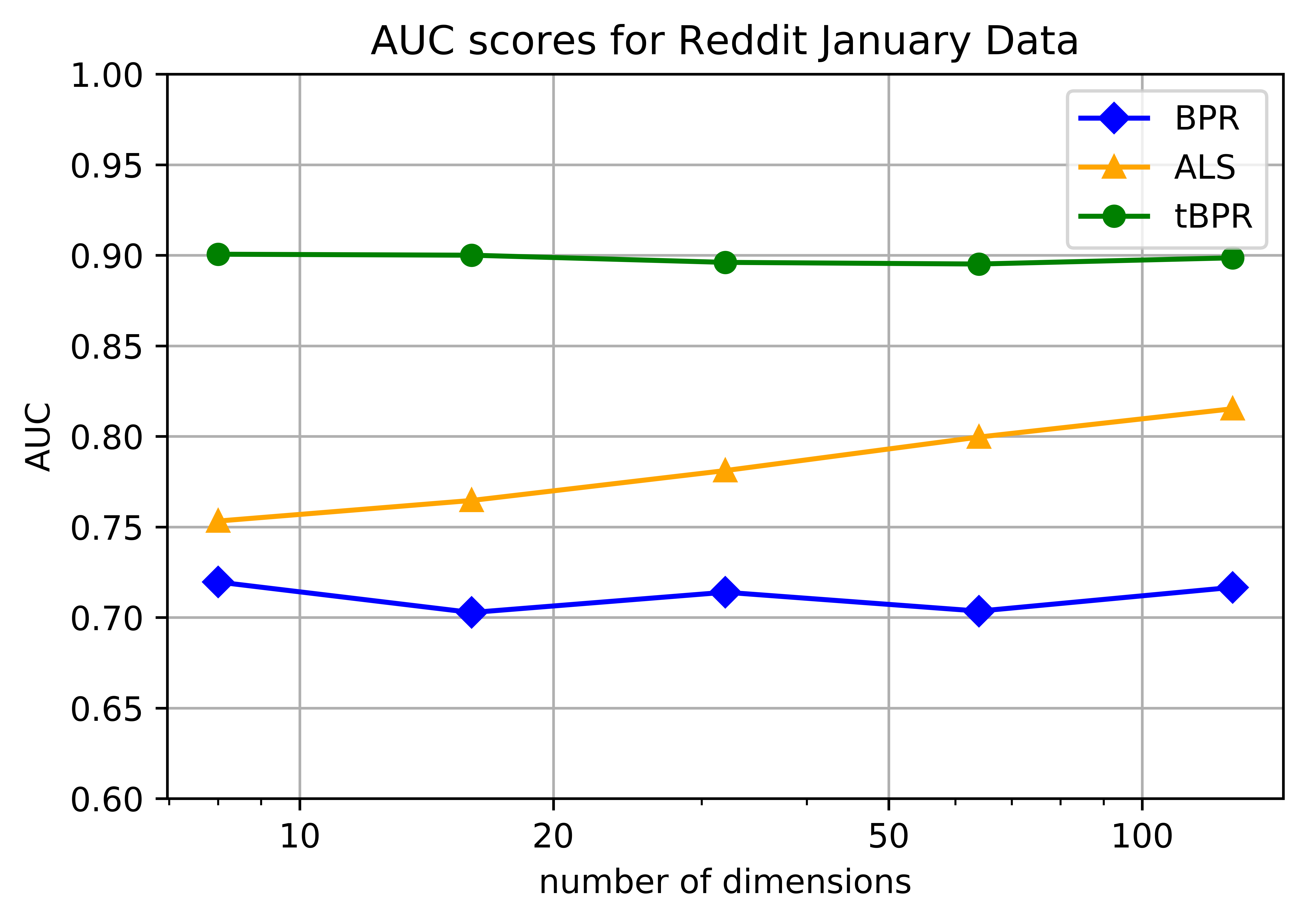}}
\caption{AUC results comparing Textual BPR vs other methods}
\label{aucs}
\end{figure}

We compare our ALS, BPR and Textual BPR models using the AUC methodology. We observe in Figure \ref{aucs} that Textual BPR performs better than the other two methods by a significant margin. We ran our experiments building vectors for various dimensions between 8 and 128. We hypothesize that this is due to the word embeddings of comments providing our model with context about the user and the subreddit. 

The best AUC scores along with the dimensions that gave the best results are in given in the table \ref{auctable}.

\begin{table}[]
\begin{centering}
\begin{tabular}{|l|l|l|}
\hline
Model    & Best AUC score & Number of Dimensions \\ \hline
MF - ALS & 0.815          & 128                  \\ \hline
BPR      & 0.717          & 32                   \\ \hline
\textbf{t-BPR}    & \textbf{0.901}          & \textbf{16}                   \\ \hline 
\end{tabular}

\caption{AUC scores for best performing models}
\label{auctable}
\end{centering}
\end{table}

\subsection{Qualitative Analysis}
We also present an example of our system's recommendation for a sample user, u/TallnFrosty. This user's history suggests an interest in sports, especially basketball, soccer and football. We find that our system's recommendations are a good mix of niche and generic subreddits which may interest him/her.
We also give an example of subreddits which are most similar to the Game of Thrones subreddit, and again find other TV shows ideologically similar to Game of Thrones in the results, as expected.

After a thorough analysis from our results, we find that the majority of our recommendations do not include reddits most popular subreddits, which shows that our model understands the actual meaning using comments, and is able to recommend novel recommends to a fair extent.

\begin{figure}[htbp]
\centerline{\includegraphics[scale=1]{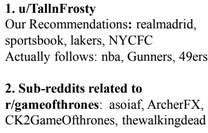}}
\caption{Examples of Textual BPR}
\label{fig}
\end{figure}

\section*{Conclusion}

Our work presents a new framework for recommending subreddits of interest for existing Reddit users based on their past interaction with the website. We present Textual BPR, an embedding based BPR system which performs significantly better than both the pre-existing real world Reddit recommender as well as tradition ALS and BPR approaches. We observe our system performs well in practice, often suggesting novel and diverse subreddits of interest. 

\section*{Acknowledgements}

We would like to thank Professor James Caverlee and Jianling Wang for their invaluable contributions and feedback for this project. We would also like to thank our classmates who gave us valuable suggestions during the initials stages as well as their critical review during our poster presentation.

\section*{Future Work}

In the future, we would like to investigate on improving serendipity, user novelty and coverage in our recommender system based on work being done in this field\cite{b5}. 

In addition, as suggested by Professor Caverlee, our item space involves polarized subreddits. User emotion in different subreddits is highly dependent on the type of subreddit. For example, subreddits involving politics are more polarized and we want to consider these latent factors using Natural Language Processsing to improve our recommender.

Our dataset also provides a controversiality score, a measure of how controversial a comment is. We plan to add this score with polarization to draw further inferences on how trolls function on sites like Reddit, and how they affect users and discussions. This would be extremely valuable as trolls are difficult to combat in real-world scenarios, especially on popular sites like Reddit.

\vspace{12pt}


\begin{thebibliography}{00}
\bibitem{b1} Steffen Rendle, Christoph Freudenthaler, Zeno Gantner, and Lars Schmidt-Thieme. 2009. BPR: Bayesian personalized ranking from implicit feedback. In Proceedings of the Twenty-Fifth Conference on Uncertainty in Artificial Intelligence (UAI '09). AUAI Press, Arlington, Virginia, United States, 452-461. 
\bibitem{b2} Ruining He and Julian McAuley. 2016. VBPR: visual Bayesian Personalized Ranking from implicit feedback. In Proceedings of the Thirtieth AAAI Conference on Artificial Intelligence (AAAI'16). AAAI Press 144-150.
\bibitem{b3} Lu, Linyuan, Matús Medo, Chi Ho Yeung, Yi-Cheng Zhang, Zi-Ke Zhang and Tao Zhou. “Recommender Systems.” CoRR abs/1202.1112 (2012): n. pag.
\bibitem{b4} t-SNE: https://en.wikipedia.org/wiki/T-distributed\_stochastic\_neighbor\_embedding.
\bibitem{b5}Mouzhi Ge, Carla Delgado-Battenfeld, and Dietmar Jannach. 2010. Beyond accuracy: evaluating recommender systems by coverage and serendipity. In Proceedings of the fourth ACM conference on Recommender systems (RecSys '10). ACM, New York, NY, USA, 257-260. DOI: https://doi.org/10.1145/1864708.1864761 
\bibitem{b6} http://aakashjapi.com/recommending-subreddits-by-computing-user-similarity-an-introduction-to-machine-learning/
\end{thebibliography}
\end{document}